\documentclass{article}

\usepackage{PRIMEarxiv}

\usepackage[utf8]{inputenc} 
\usepackage[T1]{fontenc}    
\usepackage{hyperref}       
\usepackage{url}            
\usepackage{booktabs}       
\usepackage{amsfonts}       
\usepackage{nicefrac}       
\usepackage{microtype}      
\usepackage{lipsum}
\usepackage{fancyhdr}       
\usepackage{graphicx}       
\graphicspath{{media/}}     

\usepackage{pgf-pie}  

\usepackage{tikz}
\usetikzlibrary{trees}
\usepackage{float}
\title{What We Do Not Know:\\ GPT Use in Business and Management}
\author{
  Tammy Mackenzie \\
  The Aula Fellowship for AI Science, Tech, and Policy \\
  Montreal, Canada \\
  \texttt{Tammy@theaulafellowship.org} \\
  \And
  Branislav Radeljić \\
  The Aula Fellowship for AI Science, Tech, and Policy \\
  London, United Kingdom\\
  \texttt{Branislav.Radeljic@gmail.com} \\
  \And
  Leslie Salgado \\
  University of Calgary \\
  Calgary, Canada \\
  \texttt{EvelinSalgado@gmail.com} \\
  \And
  Animesh Paul \\
  University of Georgia \\
  Georgia, USA \\
  \texttt{AP45579@uga.edu} \\
  \And
  Rubaina Khan \\
  University of Toronto \\
  Toronto, Canada\\  \texttt{Rubaina.Khan@mail.utoronto.ca} \\
  \And
  Aizhan Tursunbayeva \\
  University of Naples Parthenope \\
  Naples, Italy \\  \texttt{A.Tursunbayeva@uniparthenope.it}
   \\
  \And
  Natalie Perez \\
  Private Corporation \\
  Hawai'i, USA \\
  \texttt{Natalie.Perez@hawaii.edu}\\
  \And
  Sreyoshi Bhaduri \\
  ThatStatsGirl \\
  New York, USA \\
  \texttt{SreyoshiBhaduri@gmail.com}
}

\begin{document}
\maketitle

\begin{abstract}
This systematic review examines peer-reviewed studies on application of GPT in business management, revealing significant knowledge gaps. Despite identifying interesting research directions (best practices, benchmarking, performance comparisons, social impacts), our analysis yields only 42 relevant studies for the 22 months since its release. There are so few studies looking at a particular sector or subfield that management researchers, business consultants, policymakers, and journalists do not yet have enough information to make well-founded statements on how GPT is being used in businesses. The primary contribution of this paper is a call to action for further research. We provide a description of current research and identify knowledge gaps on the use of GPT in business. We cover the management sub-fields of finance, marketing, human resources, strategy, operations, production, and analytics, excluding retail and sales. We discuss gaps in knowledge of GPT’s potential consequences on employment, productivity, environmental costs, oppression, and small businesses. We propose how management consultants and the media can help fill those gaps. We call for practical work on business control systems as they relate to existing and foreseeable AI-related business challenges. This work may be of interest to managers, to management researchers, and to people working on AI in society.
\end{abstract}


\section{Introduction}
Decision-makers such as business leaders and policymakers face a series of hard questions on the emergence and wide diffusion of artificial intelligence tools in businesses. Management scholars have called for more research. This study specifically considers “GPT,” since its public release in November 2022 has made the tool widely accessible and available to a large populous. GPT is the commercial product of a company called OpenAI, and its name is an acronym for “Generative Pretrained Transformer.” It is a large language model (LLM) AI system, including the ChatGPT series and GPT interfaces for making applications (collectively, “GPT”). Among other things, it can make complete sentences, produce reports, respond to questions, provide images and video, provide translations, and assist in coding and other data analysis. It also makes mistakes and has no self-awareness. However, as rightly assessed by Sliż (2024: 4)\cite{Sliz2024}, “[d]espite a substantial increase in publications on GPT, research avenues exploring its implementation potential in management methodologies like BPM (business process management) remain underexplored, prompting the need for focused investigation.” Addressing this underexplored research area could offer transformative advancements in process management.\\

Management science is a social science, which relies on trustworthy information being transmitted with high fidelity \cite{Popper1934}. While relying on honor and bravery, it also provides for checks and balances on the (in)directly involved actors and their policy preferences \cite{Sagan1997}. Accordingly, peer-reviewed studies in management can have a variety of study designs, such as case-study, action research, surveys, ethnographies, and reviews or position papers \cite{Creswell2021,Dresch2015}. While position papers consider the available evidence to give recommendations, empirical papers engage with different phenomena with the goal of shedding light on potential effects. \\

This review seeks to provide a greater insight into GPT- and LLM-related business challenges, including the effect of AI tools on employment, the environmental costs of AI use, and the cultural discrimination embedded in AI systems. Knowing more about these aspects is also helpful to consultants because of their remit in the handling of management challenges \cite{Sturdy2022}. More worrisome, some unscrupulous people may make unsubstantiated claims \cite{Bouwmeester2023}. Therefore, this research aims to assist consultants who do evidence-based work and empower people working in trade and scholarly research to collaborate on researching AI in business.\\

This paper is of interest to managers, consultants, policymakers, researchers and journalists interested in the impact of GPT on business, for example in finance, marketing, human resources, and so on. Business managers and public decision-makers whom management researchers inform, and without whom management science would arguably be pointless \cite{Drucker1955}, face a lot of hard questions about technology and AI tools in general. Society is engaged in the project of AI as a socio-technical infrastructure. Understanding the challenges of AI in businesses is key to improving practices and preparing for further change. As such, this research is conducted and presented to be useful to people from industry and academia, and invites consultants and the media to collaborate with scholars in building our common knowledge.\\

\section{The State of the Debate}
\label{sec:headings}
By many accounts, we are facing a crisis in management, through abuse of power or authority \cite{DeHaan2014} in relation to 1/the standard of living and life expectancy of money \cite{Verdegem2021}, 2/ the widespread impression that some view money as more important than other people \cite{Awuah2024}, or 3/ the tendency of some business leaders to use their businesses as cover for doing harm \cite{Gottschalk2022}.\\

Drucker’s canonical contribution to the field of management science \cite{Drucker1955}, both in terms of research methods or applicability of management, remains pertinent to today’s challenges, most notably that the focus of the firm must be on the creation of customer value and, because of the impression that “culture eats strategy for breakfast,” one has to be a part of the firm to adequately guide it. Moreover, Drucker was also concerned about how time affects the quality of the information we can garner from social science, joining others \cite{Creswell2021, Popper1934} in affirming that management science is necessarily a reflective, not a predictive practice. Though well-rebutted by \cite{Denning2018}, Drucker’s critics say that his theories may have been too unspecific, inapplicable, or based on an overly rational view of the world. We decline to join that debate. At the same time, these factors behoove us to humility in asking questions that are appropriate to the tools and context, and inform the scope and discussion of the present study.\\

In brief, drawing from the above-mentioned scholars, the following characteristics are assumed of peer-reviewed scholarly research: (a) good will and intent of the authors; (b) study design that is appropriate to the question; (c) highest consideration of the latest ethical standards; (d) use of accurate data and possibility to reproduce the study; and (e) subject-matter specialists have verified and had the opportunity to criticize the claims, which can give rise to revisions. On the other hand, the nascent field of AI has been involved in some scientific controversies and potential pitfalls, relating to pertinence \cite{Chesterman2021}, inapplicability \cite{Kerner2020}, researcher bias \cite{Schwartz2022}, data manipulation \cite{Dugan2024}, hasty conclusions \cite{Bengio2023}, the influence of funders and politics \cite{Bordelon2023}, self-serving or otherwise biased conclusions \cite{Kahneman2011, Kiger2024}, but also plagiarism \cite{Kwon2024}, papers-for-money (Liverpool, 2023), and faked or unsupervised GPT outputs \cite{Kanna2024, Taloni2023}. Because of all this, each paper included in this review has been checked in detail. \\

In AI, decision-making is complicated by factors such as the rapid changes in the tech and regulatory environments, the relatively slow process of doing science, and the interdisciplinary nature of many of the issues under consideration. In the words of Hendricks \cite{Hendriksen2023}, “OpenAI’s release of GPT to the public is a paradigm shift for AI in businesses: Previous iterations of different AI systems have revolved around machine learning (ML) and big data analytics…It is no longer an exaggeration to say that this perspective was obsolete on the morning of December 1, 2022.” The short time that has passed since GPT’s release explains the shortage of scientific research on how GPT is being used in business. Indeed, almost every study we reviewed calls for more research. Moreover and to the benefit of the research community, 38\% of studies reviewed provide details of what’s missing in their sectors of scientific interest. \\

For example, as rightly put by \cite{Hamouche2023}, “[the] limited volume of research dedicated to AI in HRD [Human Resources Development] might not help fully capture and understand how AI advances can influence HR in organizations.” This goes hand in hand with the observation that “empirical studies regarding the use and declaration of human-AI collaboration at organizational frontlines remain scarce, particularly in the fields of management and marketing” \cite{Haupt2024}. In the context of cybersecurity, calls for more research take on added urgency: “While this paper purposely employs a relatively simple approach to generate adversarial attacks (on AI summarizers of financial reports), it highlights the capabilities of modern NLP (Natural Language Processing) methods, opening possibilities for developing more potent and sophisticated adversarial techniques. Given the increasing reliance on AI-powered information processing amidst the ongoing information overload, exploring adversarial attacks and developing robust NLP models becomes a crucial area of research” \cite{Leippold2023}. \\

Understandably, the interconnected and globalized challenges of decision-making around AI are manifold. They concern trust, power, climate damage, the flow of money, as well as big feelings and big potentials (\cite{Bender2021, Bengio2023, Berthelot2024, Bhaduri2024, Dehghani2024, Gebru2022, Luccioni2024, Mackenzie2024, Ovalle2023, Ricaurte2022, UNEP2024}. Amidst these challenges, it is possible to identify three types of material circulating in the public discourse on GPT in business: peer-reviewed scientific research (or scholarly research), business research reports by business consultants (or trade research), and journalism (or popular research), including websites, blogs, etc. \cite{SheridanCollege2024}. This study does not consider the role of popular and trade research. \\
	
More specifically, Hendriksen argues that GenAI disrupts the basic assumptions of business functions \cite{Hendriksen2023}, including transaction cost economics. To illustrate his concern, he offers an example from Supply Chain Management (SCM): “As AI systems become more widespread and widely adopted at both the individual and organizational levels, it changes the rules of the game for our theoretical models, assumptions, and frameworks. The reason for this is that the theories we use in SCM  are predicated on the idea that humans and organizations are the building blocks of supply chains, and supply chains emerge as phenomena from these interactions” \cite{Hendriksen2023}. Now, in his view, we have AI that takes on both sides of interactions, changing the way that agreements are negotiated and, by extension, how markets operate. This is even more relevant given that transaction cost economics is currently the dominant market theory in management science \cite{Rindfleisch2020}. Changing it changes everything, which clearly calls for additional research. \\
	
Other scholars ask for more research and specify the types of research they believe is needed \cite{Kanbach2024}, stating: “[F]uture research endeavors should include empirical approaches, analyzing measurable effects on businesses, with e.g., companies’ performance and/or the degree of measurable business model innovation as the dependent variable of the studies. Additionally, longitudinal study designs are required to understand the evolution of GAI over time and its evolving impact on business models.” As an example, one short paper reporting on the results of a survey deserves a lot more recognition for the empirical brick it places in the foundations of what we need to understand—in particular, their implications section which can trigger fresh examinations and involve researchers from different disciplines \cite{Noy2023}.\\

In terms of the culture of managing AI, rhetoricians and social scientists have identified numerous indicators of the presence of an elitist quasi-religion, sometimes called “technosaviorism” \cite{Mahmud2015} or “technocratic populism” in the globalized tech development ecosystem. As per Lechterman’s definition \cite{Lechterman2020}, “[t]echnocratic populism uses the appeal of technical expertise to connect directly with the people, promising to run the state as a firm, while at the same time delegitimizing political opponents and demobilizing the electorate by instilling civic apathy.” This is part of a bundle of philosophies, called TESCREAL, which is an acronym for “transhumanism, extropianism, singularitarianism, cosmism, rationalism, effective altruism, and longtermism,” identified by AI \& Society scholars \cite{Gebru2024, Burrell2024}. In this philosophy, technology is seen as the solution to our most pressing social and ecological problems. The claim is that some humans, superior to others in understanding, foresight, and/or access to money, should make the decisions necessary for our collective survival, and control our collective resources \cite{Lechterman2020, Srinivasan2015, Torres2023b}. Such elite utilitarian philosophies have been linked to violent authoritarian attacks on democratic systems \cite{Arendt1958}, as well as contradictions like tax evasion for the purposes of giving to charity \cite{Brennan2024, Taiwo2022}, matrix mathematics to calculate the “life-value” of future humans \cite{Wiblin2019}, and the empowerment and enrichment of people with narcissistic and abusive personality traits \cite{Olumekor2023}. It has furthermore been likened by political philosophers to “21-st century eugenics” or racialized social-sciences \cite{Anthony2024, Torres2023a}. Despite these aspects, people espousing this belief system have also been linked to major funders in AI research \cite{McAleese2023}. Effective altruists, for example, are among the largest private funders of lobbying groups around AI \cite{Bordelon2023, Cook2024}. Consequently, a manager in AI has to take into account the world-views and philosophies of the people involved, and be attentive to who is crafting policy and who is funding and choosing the topic of research. Known as the “principal-agent problem” in management, the theoretical implication is that personal and organizational interests do not always align with the interests of society or the future. Because it underlines how people can make decisions that are not in the best interests of their employer, this also indicates that the personal choice of a decision-maker can make the difference.  With evidence in hand, we can take action and identify examples, find colleagues and supporters, and design control functions that take into account the experience of people as well as our actual goals. Furthermore, we can design these goals into policy and control systems in a company \cite{Einhorn2024}; control systems are the practices of people checking to see if things are working as designed. \\
    
In businesses, this includes things like accounting and cybersecurity systems, compliance and quality control, and many other company practices. These control systems are regularly adjusted. For example, if a group of people decide that their company should not pollute, then control functions are set in place to cease polluting. These might include training, monitoring purchases, realigning product development, managing stakeholder management, setting up appropriate accounting, and so on. In short, many if not every aspect of a business. Understanding what is happening on the ground is required, when it comes to situations that are bigger than us as individuals or small groups, such as climate change and the widespread uses of AI. 

\section{Methods}
Because of GPT’s large-scale implementation across society since its release to the public in November 2022, our first a priori assumption (A1) is: GPT is being used in businesses. Consequently, the questions to ask are: How is GPT being used in businesses? (RQ1) and How is GPT affecting business processes? (RQ2) To address these questions, we employed a range of actions. We started with a systematic review of peer-reviewed studies and peer-reviewed conference papers on the use of GPT in business management. \\

First, we searched Scopus and the Web of Science. We then examined all of the retained peer-reviewed studies to look for additional studies to include. Figure 1 uses Page’s framework \cite{Page2021} for the visualization of this systematic review process for inclusion or exclusion of a source of information. It flows in time from top to bottom: Identifying sources, Screening them, and deciding on Inclusion or Exclusion. The boxes, columns, and arrows represent the decision processes taken by the researchers, starting from the top-left (Database searches) and from the top-right (Other methods).\\

\begin{figure}
  \centering
  \includegraphics[width=18cm]{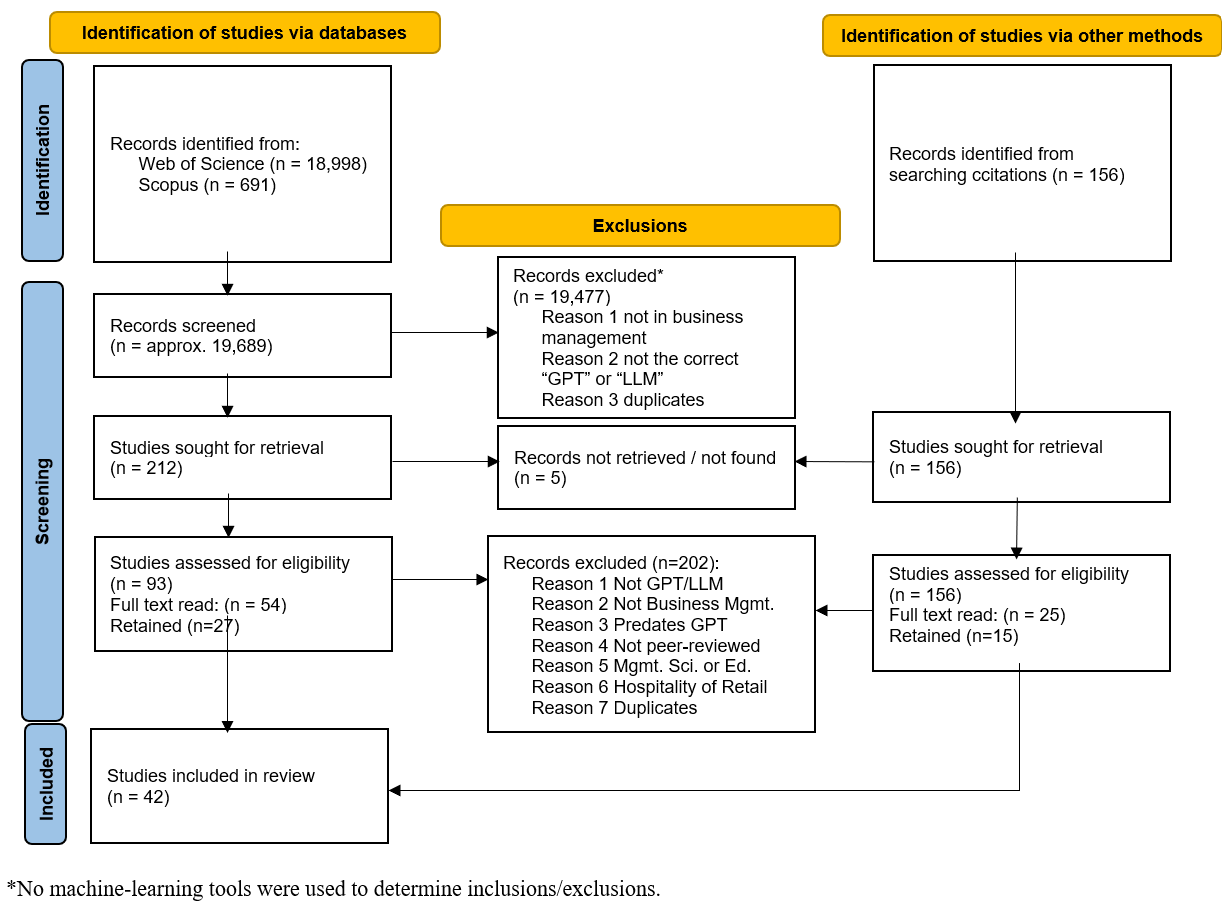}
  \caption{PRISMA Diagram of the Literature Search for Management and GPT}
  \label{fig:fig1}
\end{figure}

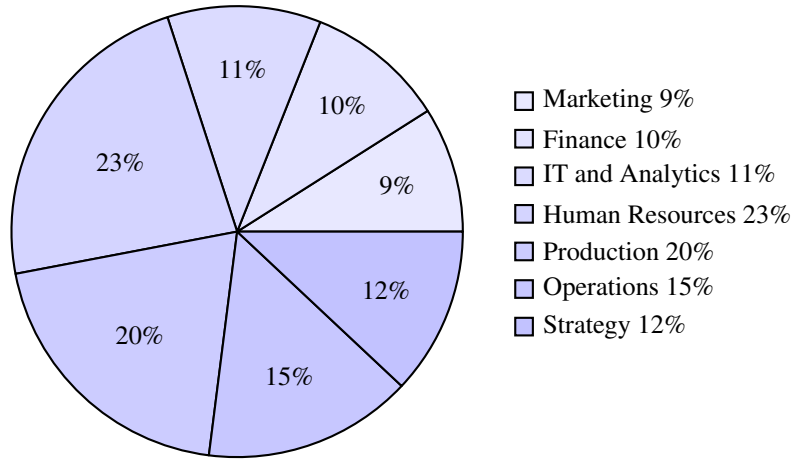
\begin{figure}[h]
    \centering
    \begin{tikzpicture}
        \pie[
            text=legend,
            radius=3,
            color={
                blue!9,
                blue!11,
                blue!14,
                blue!17,
                blue!20,
                blue!22,
                blue!24
            }
        ]{
            9/Marketing 9\%,
            10/Finance 10\%,
            11/IT and Analytics 11\%,
            23/Human Resources 23\%,
            20/Production 20\%,
            15/Operations 15\%,
            12/Strategy 12\%
        }
    \end{tikzpicture}
    \caption{Studies on GPT per management sub field}
\end{figure}

Next, we planned to proceed by conducting statistical analyses. Our study began with the premise that there would be a large number of studies, and that we would have many studies to examine to develop a grounded idea of recurrent themes within management subfields. However, given that we were able to find only 42 studies across all sub-fields (as of September 2024), a statistical analysis was not possible. To be clear, though this is not very low for a systematic review, we find that it is very low for such a widely influential topic, and fails to provide saturation on any given sub-topic, management sub-field or particular sector. \\

Studies were considered out of scope if they did not deal with business management nor with GPT, as were studies older than 2023. To keep the focus on business management, also excluded were studies in medical fields, management science, hospitality, education, and retail. We have not included consultancy reports as sources for this systematic review of peer-reviewed research, even though they are also occasionally cited in the peer-reviewed studies themselves \cite{Feuerriegel2023, FossoWamba2023, Frederico2023, Guyre2024, Haupt2024, Jackson2024, Kanbach2024}. As proposed by Ernst and Kieser \cite{Ernst2000}, “in comparison to consultants, scientists cannot simplify, they have to problematize. Their solutions have to be based on theories, while for the consultants’ solutions it suffices to be practical, to work. Scientists always have to admit that solutions derived from theories are not exact and highly contradicting, while consultants assure their clients that their solutions are definitively ‘best practices’.” Other exclusions were for using the GPT acronym for other purposes, or else for the targeted purpose, but with GPT having been a research tool and not a subject of study. Five identified studies could not be located in full-text form and so were excluded. Two studies were excluded for being of poor quality. Books that we had not already read were excluded due to the impossibility of reading them all completely. Detailed notes as to coding decisions are included in the open dataset.\\

Finally, in making coding choices and in formulating the hypotheses, we took two a priori assumptions as given: that there is a low number of studies in scope, and that there is a short time-frame for peer-reviewed research since the release of GPT in November 2022. Admittedly, there is a volume of work on AI in business done before the release of GPT. However, it has been left out of the present study. The reason for this is an assumption. The assumption is that GenAI changed the situation on the ground for businesses so drastically as to need to be investigated in isolation from previous systems. Another reason is that so few people used AI, comparatively, before the public release, that most of the studies are positional papers, many being predictive of potentialities, and ignorant of the unexpected event of the release, which was still to come. \\

The characteristics for which we coded are study type, contribution, choice of human subjects, business size, and preliminary theme (as per Gioia et al. \cite{Gioia2013}), and management subfield \cite{Fallon2024}. More precisely, the question of study type or study designs used is linked to our first hypothesis, according to which study types will be mostly positional or literature-based (H1). Regarding contribution, despite being unable to locate a robust number of studies and thus to collate their contributions in order to do a pattern analysis of commonalities and outliers, two hypotheses were key: that study contributions will mostly be calls for more research and identifying gaps (H2), and that most studies will not contribute to theory (H3). When it comes to human subjects, we are interested in precisely who is being consulted in studies reporting on how GPT is being used in management, as reflected in the hypothesis that human subject characteristics will be indicative of the value of the contribution of empirical evidence by a given study (H4). Regarding business size, we are interested in what size of businesses is being considered, which underpins the hypothesis that business size will correlate with patterns of contributions and in the findings (H5); we coded for SML small (0-50), MED medium (50-500), LRG large (500+), or MULTI multinational business. Regarding the preliminary thematic classification, although we identified common findings in the retained studies through a detailed reading of full texts, and then conducted a pattern analysis of findings to determine commonalities and outliers, as well as to further inform the discussion on gaps, it is weakly grounded as we could not reach saturation on any given topic or management subfield.\\

In terms of management subfields, here we formulated no hypothesis. However, we performed a pattern analysis to determine which fields may be more or less represented than others. We used the Business.com framework of management branches to determine the subfield categories \cite{Fallon2024}. Though other, more rigorous categorizations exist in management science \cite{Harvey2010, Mingers2015}, we chose the Business.com framework because it sufficiently and minimally categorizes the reviewed studies, enabling researchers and managers to rapidly find their subfield(s) of interest and meeting a scientific test of parsimony \cite{Popper1934}. The subfields, excluding sales, are finance, marketing, human resources, strategy, production, operations, and IT and analytics. Some studies are in more than one subfield, but have been counted in their principal subfields. Studies that cover all management subfields were listed in the strategy subfield.\\

Overall, given the small sample size, we shifted from statistical analysis to content analysis to examine the selected articles. A summative content analysis was used to assess key elements, including recurring keywords, content type, article contribution, and the presence of human subjects. This research approach was particularly useful, as the goal was to identify and quantify the frequency of specific terms, themes, or concepts, and then interpret these in relation to the broader context or content of the data \cite{Kleinheskeletal2020}. The coding scheme was developed using Zhang and Wildemuth’s (2009) eight-step process, which included defining the unit of analysis, generating a coding scheme, applying the coding scheme, checking consistency, and reporting findings. This qualitative approach allowed for a deeper exploration of the articles, crucial for addressing the research questions \cite{Zhang2009}. To ensure reliability, a panel of researchers reviewed the codes, resolved discrepancies, and agreed on the final coding scheme and results \cite{Oconner2020}.

\begin{table}[H]
\centering
\caption{Literature Reviews of ChatGPT in Business, by date of coverage}
\label{tab:literature-reviews}
\begin{tabular}{p{1.5cm}|p{3cm}|p{5cm}|p{5cm}}
\textbf{Date} & \textbf{Reference} & \textbf{Title} & \textbf{Keywords} \\
\hline
March '25 & Mackenzie et al., 2025 & What We Don’t Know: GPT use in business and management & ChatGPT; management; AI; business; SME; review; strategic management; research for impact \\
\hline
Sept '23 & Al Naqbi et al., 2024 & Enhancing Work Productivity through Generative Artificial Intelligence: A Comprehensive Literature Review & Generative artificial intelligence; work productivity enhancement; management; chatbots; ChatGPT; ethics; knowledge management; review \\
\hline
May '23 & Wach et al., 2023 & From fiction to fact: The growing role of generative AI in business and finance & Artificial intelligence (AI); generative artificial intelligence (GAI); ChatGPT; technology adoption; digital transformation; OpenAI; chatbots; technostress \\
\hline
April '23 & Frederico et al., 2023 & ChatGPT in Supply Chains: Initial Evidence of Applications and Potential Research Agenda & ChatGPT; supply chain; logistics; management; Supply Chain 4.0; industry 4.0; artificial intelligence \\
\hline
March '23 & Kanbach et al., 2024 & The GenAI is out of the bottle: generative artificial intelligence from a business model innovation perspective & Artificial intelligence; Generative AI; ChatGPT; Business model; innovation; Large language models \\
\hline
\end{tabular}
\end{table}

\vspace{-5mm}  
\section{Findings}

Table 1 presents literature reviews we identified, in chronological order of their datasets, with keywords and study contributions. For example, though notable for its rigor and width, Frederico et al.'s bold evaluation reads as follows \cite{Frederico2023}: “This viewpoint article is grounded on the few articles available in specialized magazines, blogs and company websites that approach potential applications and other issues of GPT in supply chains, as a systematic literature review was not possible due to the absence of papers approaching the subject in the research databases.” We emphasize the absence of papers. They, like others, have found very few studies pertaining to GPT \cite{Wach2023}. Consequently, facing a similar challenge, some researchers took a different approach to ours, choosing to explore the trade/business and popular media literature, to understand what is happening in the business world. We noted that many authors expressed frustration with this limitation: “Given the novelty of the topic this study can only provide a first, mainly hypothetical view on how GAI (Generative AI) will impact business models and trigger innovations along the value chain of businesses” \cite{Kanbach2024}. Sliż found that most of the articles in the 18 months following GPT’s release were about management science, not management per se; as explained, “[u]pon revisiting the bibliometric analysis in a 2024 post-research implementation, a notable research gap persists, marked by a scarcity of publications addressing both GPT implementation and business-process management (BPM). In the Web of Science database (March 2024), merely one publication was identified for the query, with a similar trend reflected in Scopus, yielding only one result (March 2024)” \cite{Sliz2024}.\\

\begin{table}[H]
\centering
\caption{Literature Reviews of ChatGPT in Business, by date of coverage}
\label{tab:literature-reviews}
\begin{tabular}{p{1.5cm}|p{3cm}|p{5cm}|p{5cm}}
\textbf{Date} & \textbf{Reference} & \textbf{Title} & \textbf{Keywords} \\
\hline
March '25 & Mackenzie et al., 2025 & What We Don’t Know: GPT use in business and management & ChatGPT; management; AI; business; SME; review; strategic management; research for impact \\
\hline
Sept '23 & Al Naqbi et al., 2024 & Enhancing Work Productivity through Generative Artificial Intelligence: A Comprehensive Literature Review & Generative artificial intelligence; work productivity enhancement; management; chatbots; ChatGPT; ethics; knowledge management; review \\
\hline
May '23 & Wach et al., 2023 & From fiction to fact: The growing role of generative AI in business and finance & Artificial intelligence (AI); generative artificial intelligence (GAI); ChatGPT; technology adoption; digital transformation; OpenAI; chatbots; technostress \\
\hline
April '23 & Frederico et al., 2023 & ChatGPT in Supply Chains: Initial Evidence of Applications and Potential Research Agenda & ChatGPT; supply chain; logistics; management; Supply Chain 4.0; industry 4.0; artificial intelligence \\
\hline
March '23 & Kanbach et al., 2024 & The GenAI is out of the bottle: generative artificial intelligence from a business model innovation perspective & Artificial intelligence; Generative AI; ChatGPT; Business model; innovation; Large language models \\
\hline
\end{tabular}
\end{table}

\vspace{-10mm}  

Table 2 presents all the retained studies on GPT in management from January 2023 to September 2024. It is sorted by study’s main methodology, and includes the study contribution, human subjects, and business size. Table 3 presents the themes covered by retained studies, with management subfields. It is sorted alphabetically by the first author’s last name. Table 4 presents the Study Types of the 42 retained studies, in three broad categories: Position/literature review (15/42), Empirical with non-human or non-business subjects (16/42), and Empirical with businesses or business people as subjects (11/42). Of the 42 studies retained on Management and GPT in the 22 months since its release, only 7 were experiments with people in a business context.\\

\setlength{\textfloatsep}{10pt}

\begin{table}[H]
\centering
\caption{Empirical Studies of ChatGPT in Business}
\label{tab:empirical-studies}
\scriptsize 
\setlength{\tabcolsep}{2pt} 
\begin{tabular}{p{2cm}|p{3cm}|p{7cm}|p{2cm}|p{2cm}}
\hline
\textbf{Study Type} & \textbf{Reference} & \textbf{Contribution} & \textbf{Human Subjects} & \textbf{Business Size} \\
\hline
Bibliometric & Isgüzar \& al, 2024 & List of authors in the field & n/a & n/a \\
\hline
Bibliometric & Hamouche \& al, 2023 & Identifies gaps & None & Not specified \\
\hline
Experiment & Noy \& Whitney, 2023 & Comparing humans on GPT-assisted / not-assisted productivity & ``Midlevel'' professionals & Not specified \\
\hline
Experiment & Barcaui \& Monat, 2023 & Comparing humans vs. GPT on project management & A project manager & SME \\
\hline
Experiment & Kernan Freire \& al, 2024 & Examining LLMs as tech support chatbots in factory production & Factory workers & Medium \\
\hline
Experiment & Venkatakrishnan \& al, 2024 & Comparing different LLMs on resume analysis & IT job seekers & Not specified \\
\hline
Experiment & Guyre, 2024 & Framework for making a management coach chatbot & Managers & Not specified \\
\hline
Experiment & Bera \& Kundu, 2024 & Methodology for GPT-enabled news analytics & Managers & Small-Medium \& Large \\
\hline
Experiment & Kernan Freire \& al, 2023 & Using GPT to enhance existing manufacturing cognitive assistance systems & Manufacturers & SME \\
\hline
Experiment & Pitkäranta \& al, 2024 & Tests a fine-tuned LLM as a customer support in finance & None & Banks / Mega \\
\hline
Experiment & Chen \& al, 2023 & Application of GPT to corporate financial statement sentiment analysis for stock market prediction & None & Large/Public \\
\hline
Experiment & Song, 2024 & Using GPT to alleviate tone manipulation in management and financial reporting & None & n/a \\
\hline
Experiment & Yu \& al, 2023 & Compare different LLMs on stock price predictions & None & Nasdaq 100 \\
\hline
Experiment & Jackson, 2024 & Used GPT to simulate logistics dynamics & None & Not specified \\
\hline
Experiment & Sainio \& al, 2024 & Custom GPT Agile Project Management problem-solving & None & Not specified \\
\hline
Experiment & Skórnóg \& al, 2023 & Measuring GPT-3.5 against ARIMA for forecasting accuracy & None & Not specified \\
\hline
Experiment & Leippold, 2023 & Uses a financial sentiment library to manipulate automated readers & None & Not specified \\
\hline
Experiment & Chuma \& de Oliveira, 2023 & Three business cases presented to GPT & None & Large/Public \\
\hline
Experiment & Naushad, 2024 & Presenting an LLM rental market analysis (no benchmarking) & None & n/a \\
\hline
Experiment & Ioanid, 2024 & Customizing GPT for Romanian business regulations & None & SME \\
\hline
Experiment & Wang \& al, 2023 & Testing ways to use GPT in manufacturing & None & Various/all \\
\hline
Experiment & Sliż, 2024 & Identifies which stage(s) of car repair service can be sped up with GPT & Specialists & Medium \\
\hline
Narrative Analysis & Newstead \& al, 2023 & How AI-generated content perpetuates gender biases in leadership development & None & Not specified \\
\hline
Position & Fosso Wamba \& al, 2023 & Survey of business GPT users & Managers & Various/all \\
\hline
Position & de Villiers \& al, 2024 & Analysis of potential AI use for sustainability reporting & None & Large/Mega \\
\hline
Position & Paul \& al, 2023 & Potential applications of GPT to consumer research & None & n/a \\
\hline
Position & Feuerriegel \& al, 2023 & Potential uses and concerns for business process management and marketing & None & Not specified \\
\hline
Position & Rivas \& Zhao, 2023 & Implications of GPT in marketing & None & Not specified \\
\hline
Position & Zhuang \& Wu, 2023 & Uses of GPT for corporate social responsibility (CSR) & None & Not specified \\
\hline
Position & Bouschery \& al, 2023 & How to use GPT for new product development & None & Various/all \\
\hline
Position & Hendriksen, 2023 & Proposes a framework for AI integration in SCM and rethinking SCM theories & None & Various/all \\
\hline
Review & Frederico \& al, 2023 & Overview of early 2023 research on GPT in supply chain management with a detailed research agenda & None & n/a \\
\hline
Review & Al Naqbi \& al, 2024 & Systematic review of GenAI research since 2017 across disciplines & None & Not specified \\
\hline
Review & Wach \& al, 2023 & Comprehensive analysis of risks associated with GPT adoption & None & Not specified \\
\hline
Review \& Content Analysis & Kanbach \& al, 2024 & How to use GPT for business model innovation & Fortune 500 employees & Large/Fortune 500 \\
\hline
Survey & Haupt \& al, 2024 & Human-AI collaboration can mitigate negative consumer responses if human control is evident & Individuals & Not specified \\
\hline
Survey & Vrontis \& al, 2023 & Connecting ChatGPT adoption to SDGs and leadership & Managers & Various/all \\
\hline
Survey & Ritala \& al, 2023 & Early survey of business users and framework for measuring implementation impact & Managers & Various/all \\
\hline
Survey & Duong, 2024 & Correlation between GPT use and entrepreneurship intent & MBA Students & n/a \\
\hline
Survey & Jo, 2024 & Implementation guidance for GPT adoption & Office workers & Various/all \\
\hline
Survey & AlQershi \& al, 2024 & How to use GPT for university management & University staff & Large \\
\hline
Survey \& Position & Talaei-Khoei \& al, 2024 & Examining how GPT integration influences firm performance & IT Executives & Medium/Large \\
\hline
\end{tabular}
\end{table}

\begin{table}[H]
\centering
\caption{Themes and subfields fields of studies on GPT in Management, January 2023 to September 2024}
\label{tab:empirical-studies}
\scriptsize 
\setlength{\tabcolsep}{2pt} 
\begin{tabular} {p{3cm}|p{1.5cm}|p{1.5cm}|p{1.5cm}|p{1.5cm}|p{1.5cm}|p{1.5cm}|p{1.5cm}|p{1.5cm}|p{1.5cm}} 
\hline
\textbf{Reference} & \textbf{Best Practices} & \textbf{Research Request} & \textbf{Benchmark/ Comparisons} & \textbf{AI Future in Business} & \textbf{Social Impacts} & \textbf{Future Analysis Frameworks} & \textbf{Firm Performance} & \textbf{Cognitive Aspects of Adoption} & \textbf{Subfield} \\
Al Naqbi \& al, 2024 & x & x & x & x & & & & & Strt \\
\hline
AlQershi \& al, 2024 & x & & & & & & & & Strt, Prod \\
\hline
Barcaui \& Monat, 2023 & & & x & & & & & & Prod \\
\hline
Bera \& Kundu, 2024 & x & & & & & & & & Fin \\
\hline
Bouschery \& al, 2023 & x & x & & x & & & & & Ops \\
\hline
Chen, \& al, 2023 & x & x & x & & & & & & Fin \\
\hline
Chuma \& de Oliveira, 2023 & & & x & & & & & & Strt \\
\hline
de Villiers \& al, 2024 & x & x & & x & & & & & Fin \\
\hline
Duong, 2024 & & & x & x & x & & & & ALL \\
\hline
Feuerriegel \& al, 2023 & x & x & x & & & & & & Mrk Prod \\
\hline
Fosso Wamba \& al, 2023 & x & & x & & x & & & & Ops \\
\hline
Frederico \& al, 2023 & x & & & x & & & & & Ops \\
\hline
Guyre, 2024 & x & & & & & & & & HR \\
\hline
Hamouche \& al, 2023 & x & x & & & & & & & HR \\
\hline
Haupt \& al, 2024 & x & & & & & & & & Mrk \\
\hline
Hendriksen, 2023 & x & x & x & x & x & & & & Ops \\
\hline
Ioanid, 2024 & x & & & x & & & & & Prod \\
\hline
Isgüzar \& al, 2024 & x & x & & x & & & & & Strt, Prod \\
\hline
Jackson, 2024 & x & x & & & & & & & Op ITA \\
\hline
Jo, 2024 & x & x & & & & & & & Prod \\
\hline
Kanbach \& al, 2024 & x & & & & x & & & & Strt \\
\hline
Kernan Freire \& al, 2023 & & & x & & & & & & Op ITA \\
\hline
Kernan Freire \& al, 2024 & x & x & & & & & & & Op ITA \\
\hline
Leippold, 2023 & x & x & x & x & x & & & x & Fin Mrk Strt \\
\hline
Naushad, 2024 & x & & & & & & & & Strt ITA \\
\hline
Newstead \& al, 2023 & x & & x & x & & & & & HR Strt \\
\hline
Noy \& Whitney, 2023 & x & x & x & & & & & & Prod \\
\hline
Paul \& al, 2023 & x & & & & & & & & Mrk \\
\hline
Pitkäranta \& al, 2024 & & & x & x & & & & & Fin Sls \\
\hline
Ritala \& al, 2023 & x & x & x & x & & x & & & HR Strt \\
\hline
Rivas \& Zhao, 2023 & x & & x & & & & & & Mrk \\
\hline
Sainio \& al, 2024 & x & & & & & & & & Prod \\
\hline
Skórnóg \& al, 2023 & & & x & & & & & & Prod \\
\hline
Sliż, 2024 & x & x & x & & & x & & & Prod \\
\hline
Song, 2024 & x & & & & & & & & Fin Strt \\
\hline
Talaei-Khoei \& al, 2024 & & & & & x & & & & Strt Prod \\
\hline
Venkatakrishnan \& al, 2024 & x & x & & & & & & & HR \\
\hline
Vrontis \& al, 2023 & x & & & x & x & x & & & Strt \\
\hline
Wach \& al, 2023 & x & x & & x & x & & & & Strt \\
\hline
Wang \& al, 2023 & x & x & x & x & & & & & Ops \\
\hline
Yu \& al, 2023 & x & x & & & & & & & Fin \\
\hline
Zhuang \& Wu, 2023 & & & & x & x & & & & Sls \\
\hline
\end{tabular}
\end{table}

\begin{table}[H]
\centering
\caption{Study Types of GPT in Management, January 2023 to September 2024}
\label{tab:study-types}
\small 
\setlength{\tabcolsep}{8pt} 
\begin{tabular}{p{6cm}|c|c}  
\hline
\textbf{Position/Literature} & \textbf{Empirical - Non-Business/No Human Subject} & \textbf{Total} \\
\hline
8 Position Papers & 13 Experiments on GPT with No Human Subjects & 15 \\
\hline
4 Reviews & 3 Surveys in Non-Business Contexts & 7 \\
\hline
2 Bibliometric Analyses & & 2 \\
\hline
1 Narrative Analysis & & 1 \\
\hline
\textbf{Total} & & \textbf{16} \\
\hline
\end{tabular}
\end{table}

Our research questions confirm that GPT is being used in businesses. However, we have insufficient scientifically collected and peer-reviewed science to make summary statements about how GPT is being used in businesses (RQ1) nor how GPT is affecting business processes (RQ2). The retained studies should be considered on a case by case basis and not in the aggregate.\\ 

Regarding our hypotheses, the first one assumes that study types will be mostly positional or literature-based, and it remains plausible. There were proportionally fewer position papers than expected, however there were only seven experiments and four surveys with people in business as subjects, indicating that only 11 of the 42 studies were in a position to provide empirical evidence of the effect of GPT in business contexts (Table 4). Many of the experiments are testing out bespoke GPT configurations, which is guidance in best practices, but they often do not loop back to benchmark or test their solution against existing solutions. This would be useful information and seems like a lost opportunity. Benchmarking GPT output to other frameworks is also a common theme for experimentation \cite{Naushad2024}. This method risks having limited utility over time, as the AI models evolve quickly and hypotheses about output quality can rapidly become obsolete. As found by others (e.g., \cite{Chen2023}), the majority of retained experiment studies are examples of GPT being tested directly, often using business problem databases. This confirms Isgüzar et al.’s  critique that “these studies only encourage businesses to use ChatGPT for instant problem-solving” \cite{Isguzar2024}. Positional papers are typically well-grounded in prior research, but it is research that does not always address GPT because it tends to pre-date it. \\
    
The second hypothesis assumes that given the low number of studies in scope and the short time-frame since the public release of GPT, study contributions will mostly call for more research in identifying gaps. This is not plausible. Much of the research retained is founded on research prior to GPT’s release. This includes empirical studies focussing on earlier applications like Chatbots or on organizations who had earlier natural language processing (NLP) systems that were not LLMs. A high proportion (75\%) of experiments concern researchers directly engaged with GPT’s output. \\

The third hypothesis assumes that given the low number of studies in scope and the short time-frame since the public release of GPT and given the crisis of reproducibility in management research, most studies will not contribute to theory. This remains plausible. There is little discussion of theory, even in the position papers, and very little theory development. Notable exceptions were Fosso Wamba et al. \cite{FossoWamba2023}, who addressed organizational learning theory, and Hendriksen \cite{Hendriksen2023}, who tackled supply chain management theory.  \\

The fourth hypothesis assumes that given the low number of studies in scope and the short time-frame since the public release of GPT, there will be fewer studies with human subjects in business contexts than without. This remains plausible. The inclusion of human subjects made 13 of the 42 studies more directly pertinent to the question of determining the real-world applications and implications of GPT use in businesses. \\

The fifth hypothesis assumes that given that business size may make a difference to results in studies examining GPTs use in business, business size will correlate with patterns of contributions and in the findings. This is inconclusive. The majority (28 / 67\%) did not identify business size or else were neutral on business size. Only four studied SMEs, and they were experiments with GPT prompting. It was not possible with such a small dataset to infer patterns of contributions.\\

Overall, the small number of peer-reviewed studies per management subfield points to gaps in our collective knowledge of how GPT is being used in business and how it is affecting outcomes. Bearing in mind that “the arrival of ChatGPT ushers in an era of vast uncertainty about the economic and labor market effects of AI technologies” \cite{Noy2023}, claims to scientific understanding about how GPT should be used and how it will affect any given business function may be premature. \\

Subfield gaps in knowledge are classified as per the management subfields they address, correlated to the number of peer-reviewed studies in our findings (Figure 2). More specifically, there is very little investigation of the environment and social impact risks raised by researchers in other fields. Confirming some other findings \cite{Frederico2023}, the retained studies neither find nor enquire on environmental questions or social displacement.  Exceptionally, one paper frankly lays out some of the principal risks to businesses from GPT use, including incorrect outputs, biases and stereotyping, copyright violations, and environmental concerns \cite{Feuerriegel2023}. \\

Regarding our aim to understand GPT- and LLM-related business challenges, it can be concluded that the effect of such AI tools on employment is understudied in peer-reviewed research. It is the focus of zero of the reviewed studies, and considered in just three \cite{Feuerriegel2023, Noy2023, Wach2023}. As nicely explained by Noy and Zhang, “[a] potent generative writing tool such as ChatGPT could conceivably either displace or augment human labor,” which further suggests that “it is not obvious whether these dynamics should be interpreted as evidence that ChatGPT will displace human workers or evidence that it will augment them.” \cite{Noy2023} \\

Experimental data for or against rhetoric on productivity and AI is understudied. This category includes just two peer-reviewed studies, both surveys, one among 275 lecturers in Malaysian public universities \cite{AlQershi2024}, and the other among 457 “college-level professionals” working on “mid-level professional writing” tasks with GPT \cite{Noy2023}. This is complemented with the discovery that there is no guidance for accounting for the high environmental costs of AI use, neither generally in management, nor specifically in accounting. There are zero studies. Peripheral to this question, one study considers GPT’s use in sustainability reporting \cite{deVilliers2024}.\\

There is almost no data to distinguish between what is happening in big enterprises and in SMEs who are using GPT. Noting that small businesses account for 70\%+ of our employment and 55\% or more of our GDP world-wide \cite{Arnold2019}, that they are essential to our ability to innovate, our manufacturing base, and as a source of entrepreneurial creativity, it is important for a lot of individual managers to understand what is going on and to determine what to expect. Into this uncertain environment, GPT-related changes for small businesses can potentially be severe. There are only 3 studies reviewed that explicitly investigate SME contexts \cite{Ioanid2024, Barcaui2023, KernanFreire2023}.\\
	
Data on the business effects of the cultural discriminations known to be embedded in AI systems \cite{Buolamwini2023, Gebru2024, Soraa2023} is insufficient, with only two examples: GPT and gender bias in leadership \cite{Newstead2023}, and GPT for corporate social responsibility reporting \cite{Zhuang2023}. This goes hand in hand with Wach et al.’s  point that \cite{Wach2023}“[t]he current literature on AI and automation does not effectively discuss the actual societal issues and worries, such as job loss and the displacement of workers.”  confirmed herein. \\

In general, the most important limitation that we are aware of is that there are so few studies that look at a particular sector or subfield in management sciences, business consultants, policy makers and journalists do not yet have enough information to make well-founded summary statements on how GPT is being used in businesses.  \\

Furthermore, few results are worse than none. Due to the law of small samples, which states that in small samples there is a higher incidence of outliers \cite{Tversky1971}, the reader is encouraged to avoid inferring any larger pattern in reality from this small group of peer-reviewed studies. At the same time, it is possible with such a small set of results to imagine what might be in the knowledge gaps, and then to go and check. While our focus on GPT has narrowed the field of potential studies, researchers should feel encouraged to look deeper into potential collaboration with the authors of studies, reviews, and their references. \\

A further limitation to our methodology is that older studies are excluded, despite having built the foundations for the studies to come. This is because we are specifically interested in GPT and it was released at the end of November 2022. However, there are a number of well-founded studies from previous years, especially on chatbots, including prior versions of GPT. These should not be excluded from other considerations, as properly elaborated by Yanxia et al. \cite{Yanxia2023}. In addition, authors working in fields other than management also occasionally publish about management and GPT use within their fields, for example when companies have been used by engineering or medical researchers as subjects or objects \cite{AlNaqbi2024}. We have not sought to compensate for this limitation. Moreover, we have excluded sales from the categories for which we measured knowledge gaps, as our exclusion of retail studies has skewed the information for the sales subfield. This limits the use of this study for people working on sales research.\\

\section{Conclusion}
As suggested by Isgüzar and colleagues, “[t]here is a need for studies that can provide insight to managers who wish to integrate GenAI as an innovative technology into their businesses” \cite{Isguzar2024}. In keeping with the principles of good science \cite{Popper1934, Sagan1997}, researchers are encouraged to collaborate to improve on other researcher’s work. Because of its nature as a general purpose technology, AI applications in business require interdisciplinary collaboration \cite{Mackenzie2024, bhaduri2024multi, dias2023using}. Furthermore, because of current changes, timely and ongoing monitoring is needed to ensure that we are aware of what we do not know. In fact, there are similar findings of large knowledge gaps on the use of GPT in the fields of journalism, K-12 education, data science, research methods, engineering education, across languages, and so on \cite{Bhaduri2024, Hays2024, Brigham2024, Khan2024, Jones2023, Mackenzie2024, KJ2024, kj2025indicmmlu, bedemariam2025potential}.\\

Easy access to AI tools such as GPT appears from public observations to be transforming management and also management education, the economy, science, and society. How this affects businesses of all sizes is disproportionately understudied and there are indications that it is rhetorically overhyped \cite{Sloane2024}. Accordingly, this paper serves to prompt researchers from science, consultancy, government, and the media to support and engage in high quality, critical scientific examination of how GPT is being used in businesses. While business consultancy and media researchers are perceived as doing a good job in the field, the challenge is to incorporate that essential front-line knowledge from consultants and journalists into rigorous frameworks from management science, so that we as researchers and as a society can begin to separate what we have hypothesized about GPT in business from what is happening in reality. We encourage special consideration for small and medium sized businesses for whom errors in AI implementation may be, proportionally to cost, more consequential \cite{McElheran2023}. \\

What may be in our gaps in research is answers about job losses and productivity, the consequences to our economy, to society, and to our planet, the consequences of biased/discriminatory systems, information on how to control for errors, and insight on many other hard, unanswered questions. We urge business professionals, like management controllers and consultants, to work with researchers to identify the elements needed to create adequate monitoring and control systems around AI use in individual sectors and in management subfields, so as to better prepare businesses for continued disruption, in immediate terms. In addition, and while it was out of the scope of this article to attempt to determine how research projects in this review were funded (and there is no indication that there is anything untoward in the studies analyzed), researchers are nevertheless cautioned to consider the influence of technocratic or TESCREAL philosophies in the field \cite{Gebru2024}. To that end and to build coalitions around important projects, we encourage scientists to be transparent about their funding and their worldview.

\bibliographystyle{unsrt}  
\bibliography{references}

\end{document}